\documentclass[3p, preprint, 12pt]{myelsarticle}

\journal{}
\usepackage{hyperref}

\usepackage{amsmath}
\usepackage{amssymb}

\usepackage{hhline}
\usepackage[english]{babel}
\usepackage[utf8]{inputenc}
\usepackage{graphicx}
\usepackage[T1]{fontenc}
\usepackage{textcomp}
\usepackage{floatflt}
\usepackage[font={footnotesize}]{caption}
\usepackage{subcaption}
\usepackage{color}
\usepackage{makecell}

\newcommand{\V}{\overline{V}}

\newcommand{\GC}{{\mathcal G}}
\newcommand{\VC}{{\mathcal V}}
\newcommand{\EC}{{\mathcal E}}
\newcommand{\GG}{\overline{G}}
\newcommand{\EG}{\overline{E}}
\newcommand{\Z}{{\mathbb Z}}

\definecolor{szin}{rgb}{0,0.44,0.4}
\definecolor{szin2}{rgb}{0.902,0.2705,0}
\definecolor{szin3}{rgb}{0,0.5,0}

\hypersetup{pdfpagemode=UseNone, pdftoolbar=true, breaklinks=true,
colorlinks=true,
linkcolor=szin2,
linktoc=page,
citecolor=szin3,bookmarks=false}
\makeindex

\newcommand{\avg}[1]{\left< #1 \right>} 
\def\aa{\alpha}

\graphicspath{{./plots/}}

\makeatletter
\def\ps@pprintTitle{%
  \let\@oddhead\@empty
  \let\@evenhead\@empty
  \let\@oddfoot\@empty
  \let\@evenfoot\@oddfoot
}
\makeatother

\begin{document}

\begin{frontmatter}

\title{Empirical study of the role of the topology in spreading on communication networks}

\author[CEUM,Renyi,SBRAS]{Alexey Medvedev\corref{correspondingauthor}}
\cortext[correspondingauthor]{Corresponding author at: Department of Mathematics and its Applications, Central European University, Nador ut. 9, Budapest, 1051, Hungary}
\ead{an\_medvedev@yahoo.com}

\author[CEUCNS,BME]{Janos Kertesz}
\ead{janos.kertesz@gmail.com}

\address[CEUM]{Department of Mathematics and its Applications, Central European University, Nador ut. 9, Budapest, 1051, Hungary}
\address[SBRAS]{Laboratory of Graph Theory, Sobolev Institute of Mathematics, Acad. Koptyug av. 4, Novosibirsk, 630090, Russia}
\address[Renyi]{Renyi Institute of Mathematics, Budapest, Reáltanoda ut. 13-15, 1053, Hungary}
\address[BME]{Technical University of Budapest, Budapest, Budafoki ut. 8, 1111, Hungary}
\address[CEUCNS]{Center of Network Science, Central European University, Nador ut. 9, Budapest, 1051, Hungary}

\begin{abstract}
Topological aspects, like community structure, and temporal activity patterns, like burstiness, have been shown to severly influence the speed of spreading in temporal networks. We study the influence of the topology on the susceptible-infected (SI) spreading on time stamped communication networks, as obtained from a dataset of mobile phone records. We consider city level networks with intra- and inter-city connections. The networks using only intra-city links are usually sparse, where the spreading depends mainly on the average degree. The inter-city links serve as bridges in spreading, speeding up considerably the process. We demonstrate the effect also on model simulations.  
\end{abstract}


\end{frontmatter}

\section{Introduction}
Spreading is one of the most important dynamic processes on complex networks \cite{Vespignani2012, RevModPhys2015} as it is the basis of broad range of phenomena from epidemic contagion to diffusion of innovations. The underlying complex network is that of human interactions, however, this ``Social Connectome'' cannot be characterized by a static graph. The links may be active for some, often only short, period, and then inactive for the rest of the time \cite{Bansal_etal2010}. An adequate framework to describe this situation is that of temporal networks \cite{HolmeSaramaki_PhyRep2011}. 

Recently much effort has been invested in understanding the main governing factors of spreading on networks \cite{Karsai_etal2011, HorvathKertesz2014, Delvenne_etal2015}. Topological aspects like community structure \cite{Fortunato2010} and temporal activity patterns like burstiness \cite{Barabasi2005} severly influence the speed of spreading \cite{IribarrenMoro2009, Karsai_etal2011, Miritello_etal2011}. 

Different approaches have been applied to tackle this problem. Considerable effort has been made to treat related models analytically and by numerical simulations \cite{Vazquez_etal2007, Liben-NovellKleinberg2008, IribarrenMoro2009, Jo_etal2014, Delvenne_etal2015, HorvathKertesz2014}. Another possibility is to use empirical data about  temporal networks and play the game of a susceptible-infected (SI) or susceptible-infected-removed (SIR) process with those data. Such data include email logs \cite{Newman_etal2002,Vazquez_etal2007}, records of face to face encounters \cite{Stehle_etal2011} or mobile call billing information \cite{Karsai_etal2011,Miritello_etal2011}. Mobile call records provide detailed insight into the dynamics of human interactions and can especially well be applied to study different aspects of communication processes \cite{Blondel_etal2015}. In this paper we will follow this line. 

The motivation of our research comes partly from the recent observation that very little change in the topology of the underlying network may have strong impact on the spreading process. In \cite{HorvathKertesz2014} dramatic differences were demonstrated between spreading curves for trees and graphs, which are only locally tree-like. For very bursty activity patterns, i.e., if the distribution of the inter-event times decays sufficiently slowly, the spreading curve for tree structure shows abrupt steps irrespective of the averaging, while for trees with ``bridge'' that result in large loops, the curves become smooth, indicating a decisive role of these large loops. This has been proven in the mean time with mathematical rigor \cite{MedvedevPete2016}.

On the other hand it was shown \cite{Pan_etal2011} that in mobile call networks subscribers from a city can be considered as a kind of a community. We decided therefore to study how the communication and the spreading of information is structured in the cities. What are the most important factors of the spreading speed in such a realistic sample? How  is it influenced by external links? We use time stamped mobile call data for simulating the SI process on the real communication networks.

The paper is organized as follows. In the following Section we describe the data and the method of analysis. In Section 3 we present the results. Then in Section 4 we show that the results can be reproduced on the model of a communication network with simple assumptions about the bursty dynamics. We conclude the paper with a discussion.

\section{Data and analysis}

The dataset contains records of mobile phone calls (MPC) of one service provider of one European country. The call records contain the caller and callee's hashed codes, the starting time of the call and it's duration. The dataset spans over a five-month period from August to December of 2007 and contains 2,271,071,378 call records between 53,955,465 users. Among these users, 9,769,376 users have active contracts with the service provider during the observed period and we call them as \textit{company users}. The rest of the users are called \textit{non-company users}. Company users may have additional information about their location, given by the ZIP code stated in the contract. The number of company users with ZIP codes is 6,272,586, others have no declared ZIP code. The data contains two types of calls: 1) between two company users and 2) the calls from a company user to a non-company user. Since the data comes from one service provider, we have no information about the initiated calls or geographical information of non-company users.

The MPC data generates a temporal network $\GC$ with the node set $\VC$ of users and the set of events $\EC$ of calls of users. The \textit{underlying network} is a projection of a temporal network $\GC$ on a static graph $\Gamma$ with the same node set $\VC$ and there is a link between $u,v\in\VC$ iff there is at least one event between $u$ and $v$ in $\EC$. We call the nodes representing company users with known ZIP code as \textit{white} nodes, the nodes representing company users without ZIP code as \textit{grey} nodes and the nodes of non-company users as \textit{black} nodes. We assume that when a call happens, spreading can go in both directions, from a callee to a caller or vice versa, thus, for the purpose of current research, the network is considered to be undirected. Since there is no data about calls between non-company users, there are no links between nodes representing these users.

\begin{figure}[h]
  \centering
  \includegraphics[scale=1]{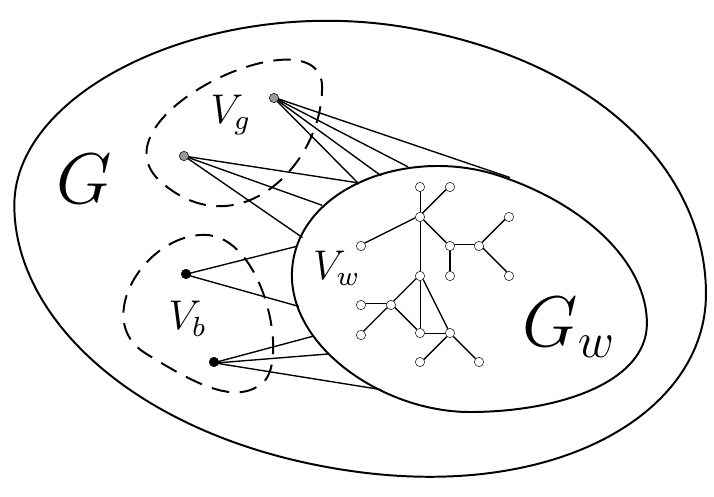}
  \caption{Schematic structure of networks $G_w$ and $G$. White nodes are denoted as $V_w$, grey nodes as $V_g$ and black nodes as $V_b$.}
  \label{fig:graph_illustration}
\end{figure}

In the current research we consider sub-networks of cities defined as follows. For each city the network $G_{w}=(V_{w}, E_{w})$ is defined as the set of white nodes $V_{w}$ with the ZIP code matching one of the ZIP codes of the city with the set of links $E_{w}$ that represent calls between those users. The network $G_w$ is an induced sub-network of $\Gamma$. The network $G=(V,E)$ is defined as an extension of network $G_w$ with the set of grey nodes $V_g$ and black nodes $V_{b}$, that have links to the nodes from $V_w$. Along with the grey and black nodes we add links connecting $V_g$ and $V_b$ to $V_w$. Clearly $G_{w}\subset G$ and $V=V_{w}\cup V_{g}\cup V_{b}$ and by construction there are no links within and between nodes from $V_{g}$ and $V_{b}$ in $G$. Schematically the networks $G_{w}$ and $G$ are represented in the Figure~\ref{fig:graph_illustration}.

The grey and black nodes of degree $k\geqslant 2$ in the network $G$ can be considered as \textit{bridges} between white nodes. We compare the spreading on the set of \textit{white} nodes \textit{without} bridges (in $G_w$) and \textit{with} bridges (in $G$). Therefore we perform the modification of each network $G$ by removing all grey and black nodes of degree 1. Then we consider the largest connected component $LCC(G_w)$ in the network $G_w$ and $LCC(G)$ in the network $G$. We denote as $\avg{k_w}$ the average degree of white nodes.

The SI spreading process on the network is defined as follows. Each node is labelled by one of the two states: susceptible ($S$, non-informed) or infected ($I$, informed). Suppose that at the time $t=0$ all nodes are in the susceptible state and one node, the \textit{initiator}, chosen at random, turns into the infected state. The infection is passed with time $t$ through links from infected nodes to susceptible ones \textit{instantaneously} at the first time when an interaction happens between these two nodes. The process runs until the connected component of the initiator becomes infected. It may happen that the data span is not enough to infect the whole network, thus periodic boundary conditions in time are applied for the data. 

We define a spreading curve $N(t)$ as the ratio of \textit{white} nodes infected in the process by time $t$ to the size of the connected component of the initiator. In our simulations we perform $M$ runs of the SI spreading process with random initiators from the largest connected component of the underlying network. The \textit{average spreading curve} is then defined as the average curve $\avg{N(t)}$, where
$$\avg{N(t)}= \frac{\sum\limits_{i=1}^{M}N_i(t)}{M}.$$
The \textit{characteristic time} is defined as the time $\tau$, such that $\avg{N(\tau)}=1/2$.

The the behaviour of SI spreading on each network $G_w$ and $G$ is studied using the corresponding average spreading curves $\avg{N_w(t)}$ and $\avg{N(t)}$. At each run $i$, where $i=1,\dots,M$, the initiator is chosen from the set of white nodes of $LCC(G_w)$ with degree at least $2$ to avoid slowdown related to possible bottlenecks at the very beginning of the process. The characteristic times of SI spreading on networks $G_w$ and $G$ are denoted as $\tau_w$ and $\tau$ correspondingly.

\section{Empirical results}\label{sec:results}

\begin{figure}[h]
  \centering
  \includegraphics[width=0.85\linewidth]{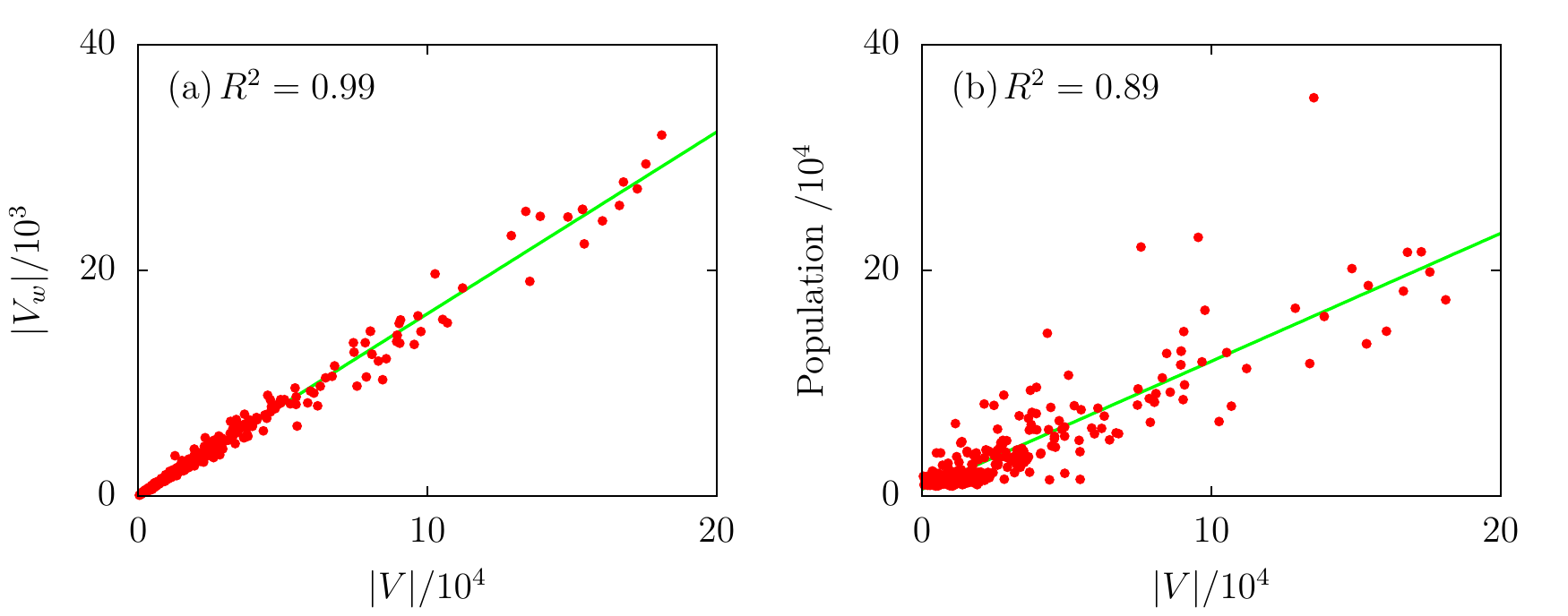}
  \caption{(a) Dependence between the size of the network $G$ and the number of white nodes $|V_w|$ for each city. (b) The relationship between the size of the network $G$ and the population for each city. The linear trend is vague, which represents the difference in coverage proportion of the company in each city.}
  \label{fig:num_pop_white}
\end{figure}

We consider the sample of $N\simeq 300$ cities with population between 10,000 and 300,000 people. Cities of smaller population are not considered because of small size of produced networks, and cities with larger population are underrepresented in the data. For each city $i$, where $i=1,\dots, N$, we construct networks $G_{w}(i)$ and $G(i)$. The size of networks $G$ is proportional to the number of white nodes (see Figure~\ref{fig:num_pop_white}, (a)), however the dependence of the size on the population of cities is vague (see Figure~\ref{fig:num_pop_white}, (b)), which is explained by differences in the coverage of the company in different cities. Thus, it follows from the data that the number of bridges, or the total number of grey and black nodes $|V_g|+|V_b|$, is also proportional to the size of the networks $G$, such that 
\begin{equation}\label{eq:proportion}
|V_g|+|V_b|\simeq 5|V_w|,
\end{equation}
which is further used in the Section~\ref{sec:modelling}. 

\begin{figure}[h]
  \centering
  \includegraphics[scale=0.75]{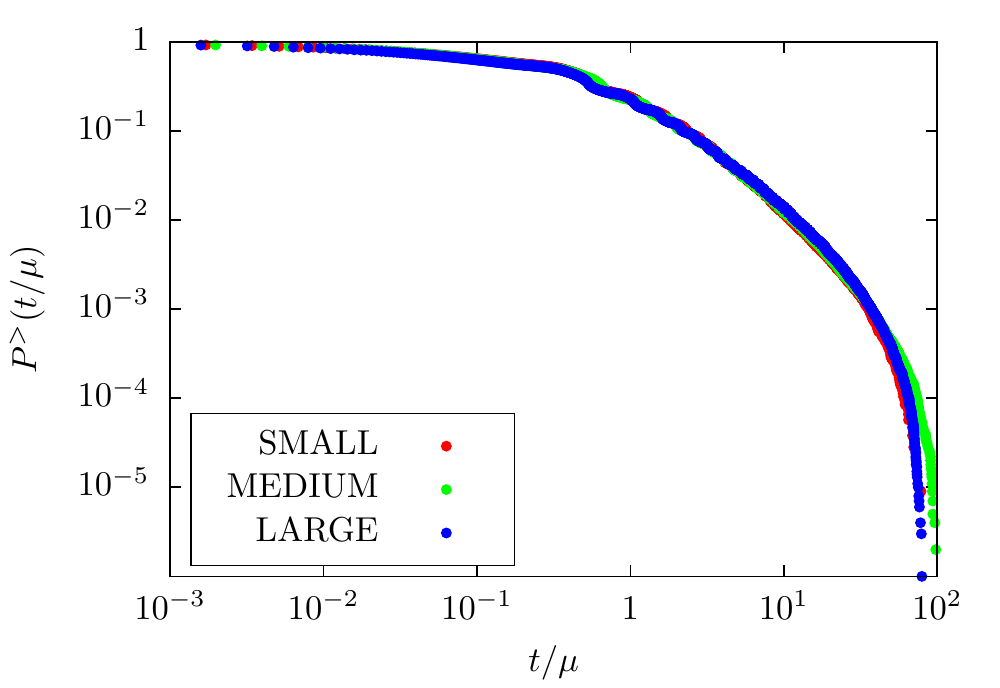}
  \caption{The inter-event time distributions of three cities of different sizes of $V_w$. A small sized city has $|LCC(G_w)|\sim 10^3$, a medium sized city has $|LCC(G_w)|\sim 10^4$ and a large is with $|LCC(G_w)|\sim 10^5$.}
  \label{fig:interevent_time}
\end{figure}

The average inter-event time for each city $i$, where $1\leqslant i\leqslant N$, is denoted as $\mu_i$ and the corresponding standard deviation is denoted as $\sigma_i$. These values are very similar for all cities and $\mu_i\in(0.04, 0.08)$ and $\sigma_i\in(0.1, 0.2)$ for almost all cities $i$. The burstiness coefficients $B_i$ belong to the interval $(0.41,0.46)$, indicating the bursty interaction patterns in all cities. The tail distributions $P^{>}(t/\mu)=\int\limits_{t}^{\infty}p(s/\mu)ds$ of inter-event times in the cities are found to be similar for cities of different population size (see Figure~\ref{fig:interevent_time}), which is explained as that people follow same patterns of calls irrespective of the size of their community.

\begin{figure}[h]
\centering
\begin{minipage}[b]{0.49\textwidth}
\includegraphics[scale=0.61]{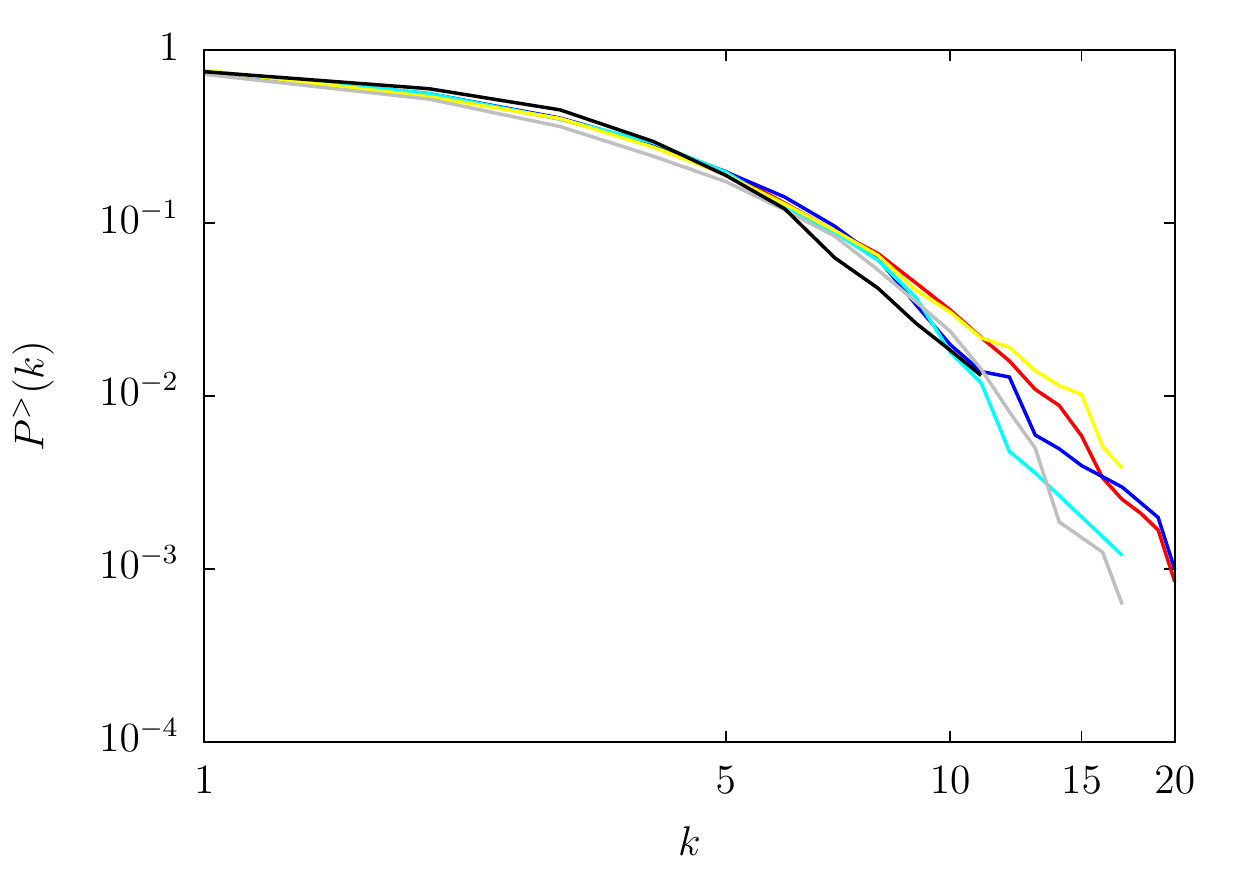}
\caption*{(a) Degree distributions}
\end{minipage}\hfill
\begin{minipage}[b]{0.49\textwidth}
\centering
\includegraphics[scale=0.61]{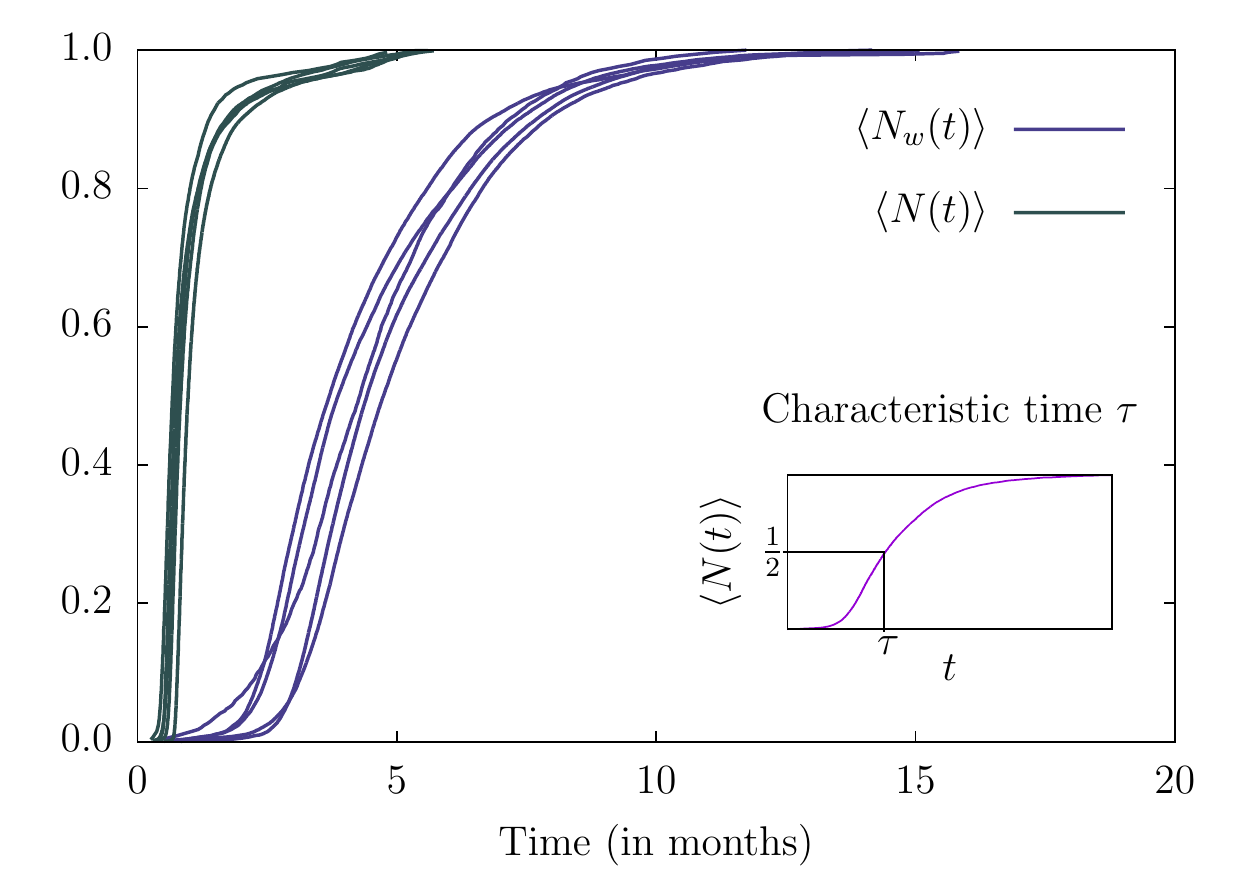}
\caption*{(b) Spreading curves}
\end{minipage}
\caption{(a) Tail distributions of node degrees and (b) average spreading curves $\avg{N_w(t)}$ (dark blue) for 6 typical networks of cities $G_w$ with average degree $\avg{k_w}=3\pm 0.2$ are given. Respective average spreading curves $\avg{N(t)}$ for networks $G$ of same cities are also given in dark grey in (b). The inset of (b) illustrates the definition of the characteristic time $\tau$.}
  \label{fig:curves_degree}
\end{figure}

We perform $M=100$ runs of SI spreading process on each of the networks $G_{w}$ and $G$ and observe that the cities with similar average degrees $\avg{k_w}$ have similar degree distributions. On the Figure~\ref{fig:curves_degree}, (a), a sample of tail distributions of node degrees $P^{>}(k)=\sum_{k}^{\infty}p(s)ds$ is presented. It is also seen that spreading curves have the same shape for the cities with similar average degrees. The sample of average spreading curves for the same cities with $\avg{k_w}=3\pm 0.2$ is given on the Figure~\ref{fig:curves_degree}, (b), along with the illustration of the notion of characteristic time. 

\begin{figure}[ht]
\centering
\includegraphics[scale=0.625]{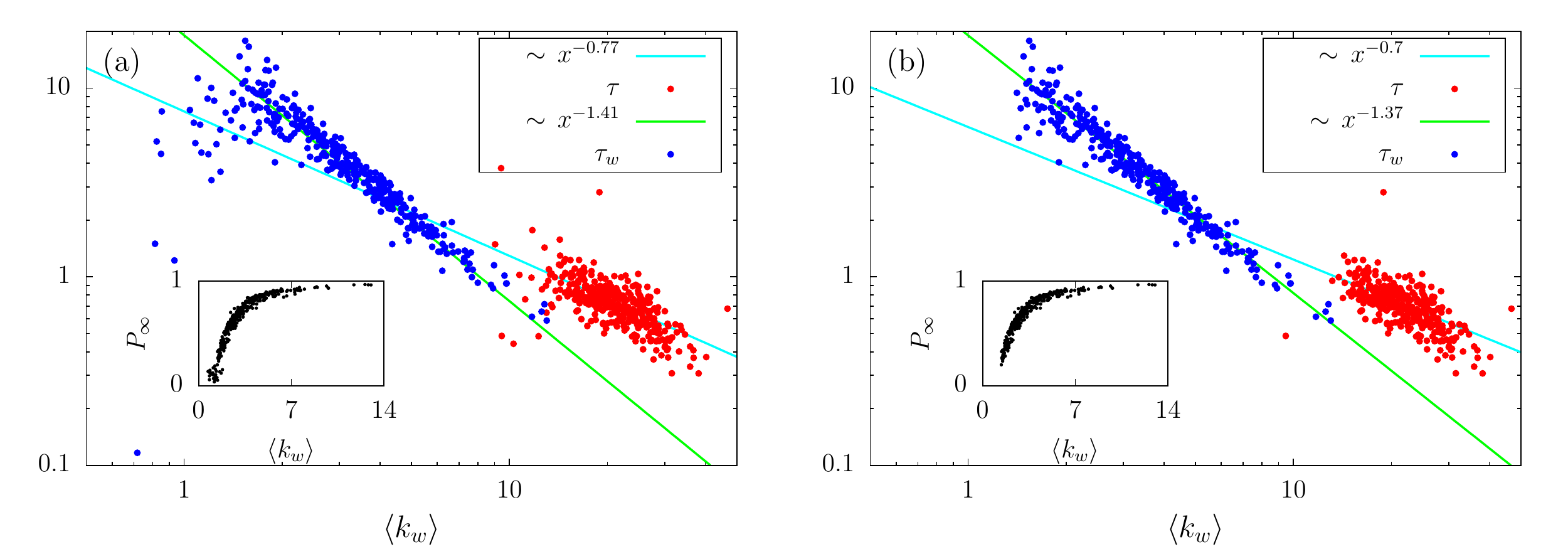}
\caption{The characteristic times $\tau_w$ (blue dots) and $\tau$ (red dots) relative to average degree of white nodes $\avg{k_w}$ in the networks $G_w$ and $G$. The insets show the dependence of $P_{\infty}$ on the average degree of white nodes $\avg{k_w}$. (a) The whole sample of cities is plotted, with large scattering around $\avg{k_w}\sim 1$. In this case the networks become very sparse and the size of $LCC$ is comparatively small. (b) The same plot is obtained by deleting those cities with $P_{\infty}<20\%$, which gives clearer results. }
\label{fig:avdeg_tau_both}
\end{figure}

In agreement with our expectations we see a radical decrease of the characteristic times, when we switch from $G_w$ to $G$ (Figure~\ref{fig:avdeg_tau_both}). Remember that $P_{\infty}$ is the relative size of the $LCC(G_w)$ to the total size of $G_w$. Then on the insets of Figure~\ref{fig:avdeg_tau_both} we see that the dependence of $\avg{k_w}$ over the $P_{\infty}$ shows the typical percolation behaviour. We observe when $P_{\infty}<20\%$, the networks show scattered results in characteristic time (see Figure~\ref{fig:avdeg_tau_both}, (a)), thus we perform a truncation of those cities and present the results on Figure~\ref{fig:avdeg_tau_both}, (b). 

We find a clear power-law dependence of the characteristic time $\tau_w$ on the average degree $\avg{k_w}$ in networks $G_w$ (see Figure~\ref{fig:avdeg_tau_both}, blue plots). At first sight it is surprising that $\avg{k_w}$ and not the size of the $LCC(G_w)$ controls the process. This may be explained by the small size effect of networks $G_w$, since most of the sizes of $LCC(G_w)$ belong to the interval $[10^2, 10^4]$.

The introduction of bridges of grey and black nodes drastically decreases the characteristic time (see Figure~\ref{fig:avdeg_tau_both}, red plots). However, the dependence of the characteristic time $\tau$ on the average degree $\avg{k_w}$ in networks $G$ is less apparent. The reason the characteristic time is so much decreased ($\tau\in[0.5;1.5]$) is that spreading curves start to look almost the same for all values of $\avg{k_w}$, thus smearing the effect of the topology of the underlying network. In the next Section we model this effect on the null network models.

\section{Modeling}\label{sec:modelling}

\subsection{Model}

In this Section we consider several network models, which we describe here briefly. Two of them are basic network models: the Barab\'asi-Albert $(BA)$ \citep{BarabasiAlbert1999} and the Erd\H os-R\'enyi $(ER)$ \citep{ErdosRenyi1959}. The infinite square lattice with next nearest neighbour interactions (NNN) is defined as the infinite regular two-dimensional grid of nodes $[i,j]$, where $i,j\in\Z$, and nodes $[i_1,j_1]$ and $[i_2, j_2]$ are connected with an edge, if $|i_1-i_2|+|j_1-j_2|=1$ or $|i_1-i_2|+|j_1-j_2|=2$. The two-dimensional \textit{finite} square lattice with NNN $(Z^2)$ is defined as an induced sub-network of an infinite lattice on $n^2$ nodes $V(Z^2)=\{[i,j]:1\leqslant i,j\leqslant n\}$. The last considered model is a network model $K$ designed to reflect the community structure of the society, developed by Kumpula et.al. in \cite{KSKK07}. The important properties of all four above models are summarized in the Table~\ref{tab:network_properties}. Degree distributions of networks $Z^2$, $ER$ and $BA$ are known, when the degree distribution of network $K$ can be well described by the log-normal distribution and it is stable with respect to the size of the network.

\begin{table}[ht]
\centering
\begin{tabular}{|c|c|c|c|c|}
 \hline   & \makecell{Degree \\ distribution} & Small-world & \makecell{Percolation \\ threshold} & Communities\\ 
 \hline  $Z^2$ & $Constant$ & - & + & - \\ 
 \hline 
 $ER$ & $Poisson$ & + & + & - \\ 
 \hline 
 $BA$ & $Power-law$ & + & - & - \\ 
 \hline 
 $K$ & $Log-normal$ & + & + & + \\ 
 \hline 
\end{tabular}  
\caption{Main properties of the considered network models. }
\label{tab:network_properties}
\end{table}

We construct a simple model of networks $G_w$ and $G$ that replicates the observed behaviour of the SI spreading process. Our model is based on the observation that considered networks $G_w$ only give a partial insight into the whole network of calls in the city, thus not all connections between nodes are revealed. We also note from the data that the number of white nodes is proportional to the total number of bridges, which is given by \eqref{eq:proportion}.

Let us describe the model. Consider a network $\GG=(\V,\EG)$ on $n$ nodes represented by a network model with the average degree $d$. We define a \textit{diluted} network $\GG^p_{w}$ as the initial network $\GG$ with each link being deleted with probability $p$, where $0< p<1$. Define the nodes of the network $\GG^p_{w}$ to be white nodes. For each $\GG^p_{w}$ we define the network $\GG^p$ by adding the set of bridge nodes to $\GG^p_{w}$ and connecting each bridge node to exactly two white nodes at random. We model the interaction patterns by attaching i.i.d. random variables $\xi_i$ to the links, which represent the inter-event times. We consider two types of inter-event time distributions: 1) the \textit{power-law distribution} $pow(t_{min},\aa)$ with the following density function:
\begin{equation*}
p_{pow}(t)=
\begin{cases}
t_{min}^{\aa}\aa\frac{1}{t^{\aa+1}},& \text{if } t_{min}\leqslant t;\\
0, & \text{otherwise,}
\end{cases}
\end{equation*}
and 2) the \textit{shifted exponential distribution} $exp(t_{min},\lambda)$ with the following density function:
\begin{equation*}
p_{exp}(t)=
\begin{cases}
\frac{1}{\lambda-t_{min}}\exp\left(-\frac{t-t_{min}}{\lambda-t_{min}}\right),& \text{if } t_{min}\leqslant t;\\
0, & \text{otherwise,}
\end{cases}
\end{equation*}
The distribution $exp(t_{min},\lambda)$ is defined such that it has the same support as $pow(t_{min},\aa)$.

\subsection{Simulation}\label{sec:simulation}

The simulation is performed on networks $\GG^p_w$ and $\GG^p$ constructed from a network $\GG$, given by four network models defined earlier: 1) Finite square lattice with NNN $(Z^2)$; 2) Erd\H os-R\'enyi model $(ER)$; 3) Barab\'asi-Albert model $(BA)$; 4) the model of Kumpula et.al. $(K)$. Each considered network model has $n=5000$ nodes (network $Z^2$ has $4900$ nodes) and the parameters of these models are tuned in such way that the average degree $d$ in the original $\GG$ is around $12$. The link deletion probability $p$ goes from $0$ to $1$ with increment $0.05$. The number of bridges added to each network is $5|V_w|$, which is obtained from the data. 

We perform two sets of spreading runs, with $M=100$ runs in each set, and in each run the starting node is chosen at random from the set of white nodes in $LCC(\GG^p_w)$ of degree greater or equal than two. The first set of runs has power-law inter-event times distribution on edges with $t_{min}= 0.008$ and $\alpha=1.2$, and the second set has shifted exponential distribution with the same $t_{min}$ and the parameter $\lambda$ is chosen in the way that two distributions have same mean $\mu$.

\begin{figure}[ht]
\centering
\begin{minipage}[b]{0.49\textwidth}
\centering
\includegraphics[scale=0.55]{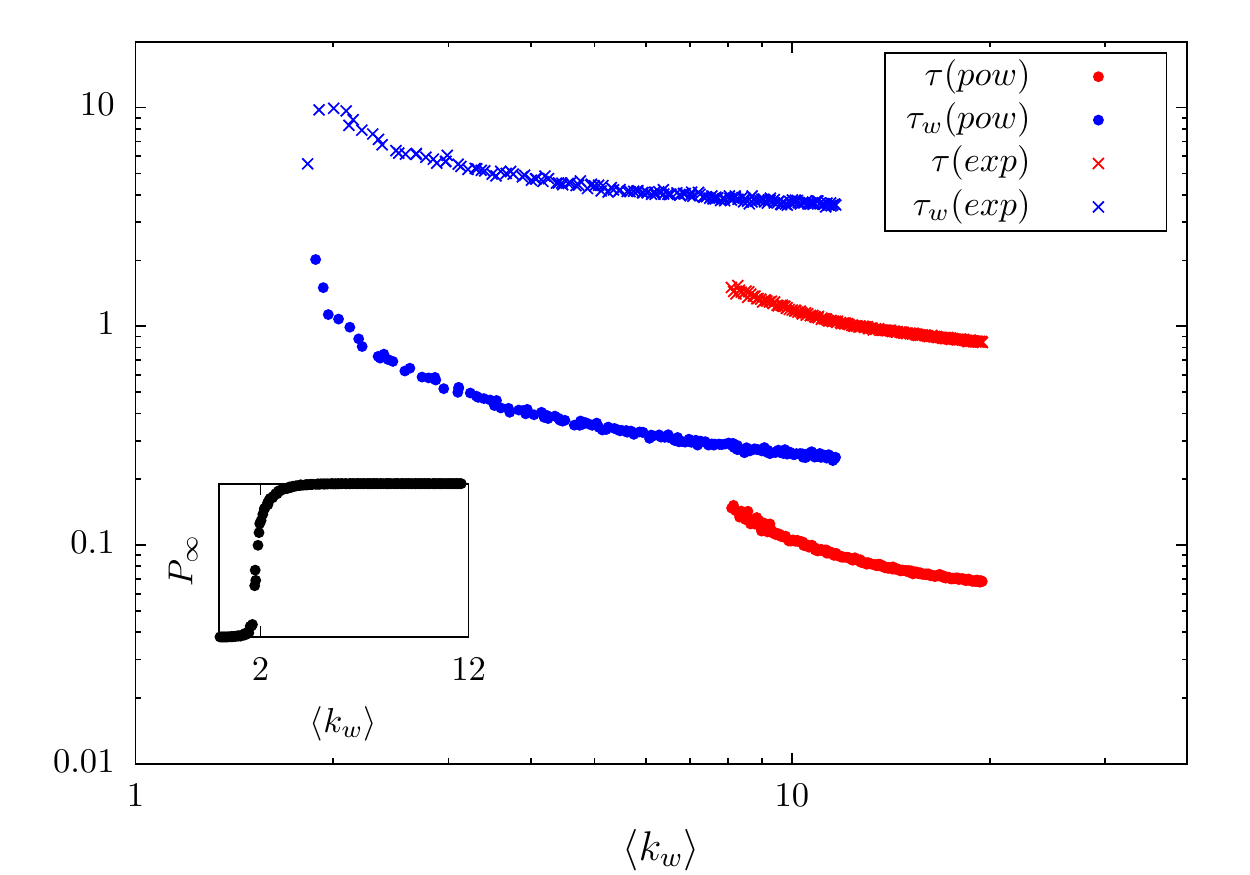}
\caption*{(a) Regular lattice with NNN}

\end{minipage}\hfill
\begin{minipage}[b]{0.49\textwidth}
\centering
\includegraphics[scale=0.55]{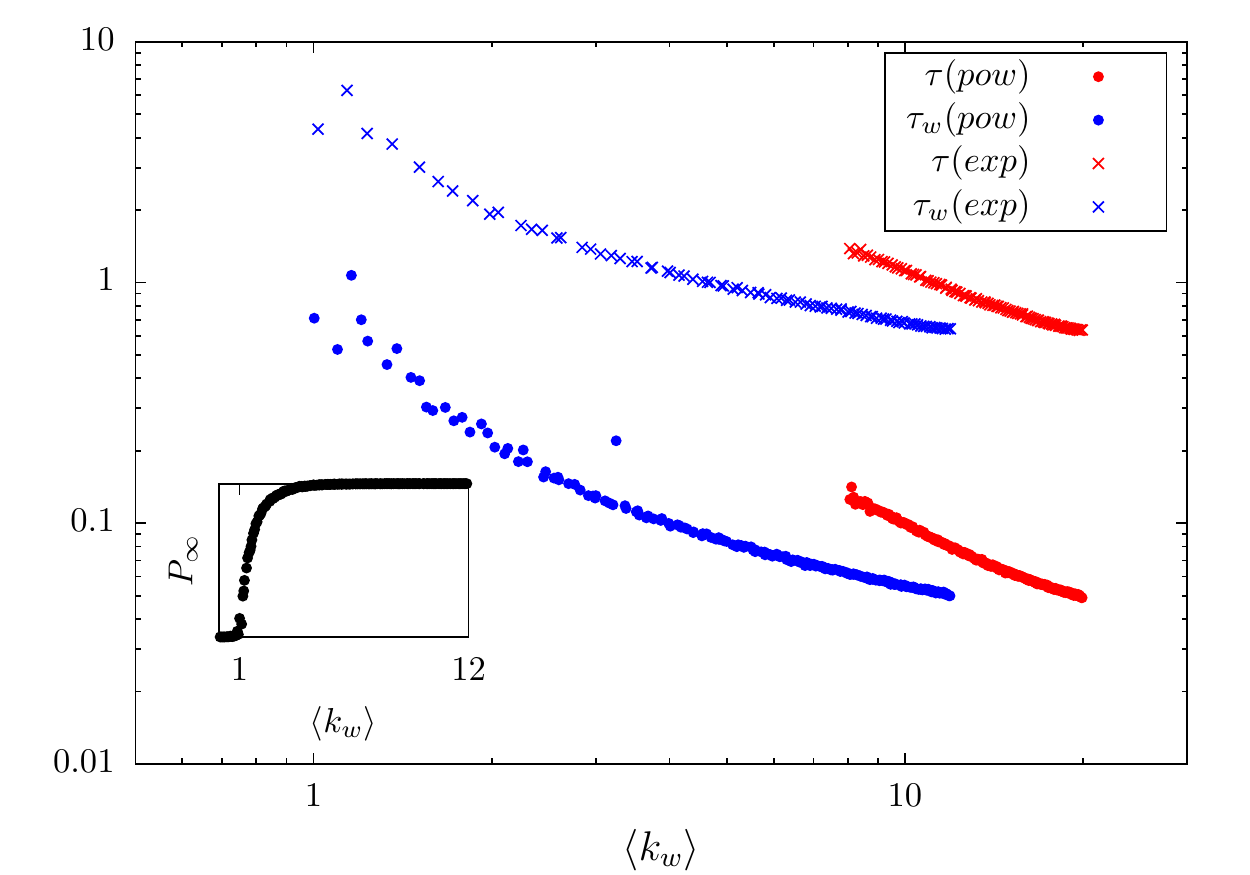}
\caption*{(b) Erd\H os-R\'enyi network}
\end{minipage}
\\

\centering
\begin{minipage}[b]{0.49\textwidth}
\centering
\includegraphics[scale=0.55]{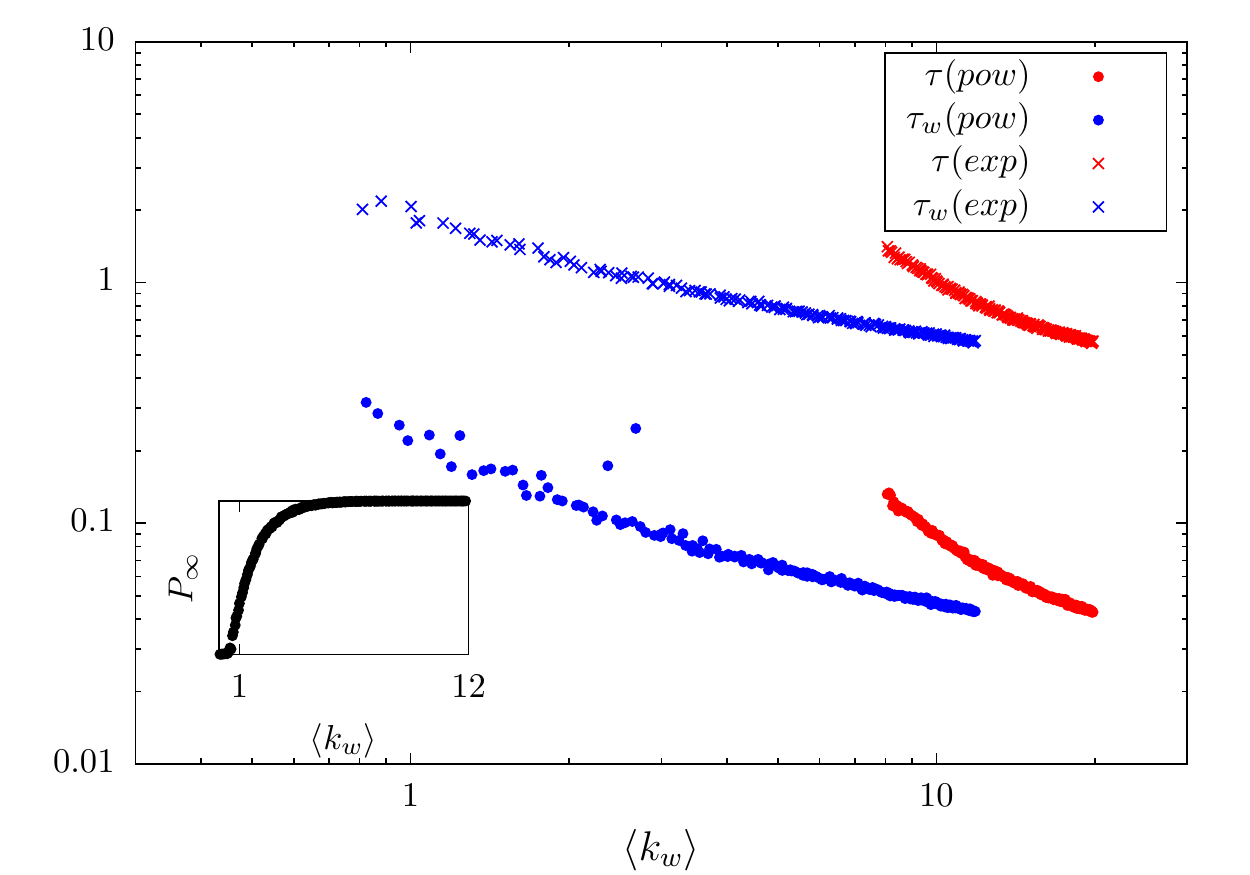}
\caption*{(c) Barab\'asi-Albert network}
\end{minipage}\hfill
\begin{minipage}[b]{0.49\textwidth}
\centering
\includegraphics[scale=0.55]{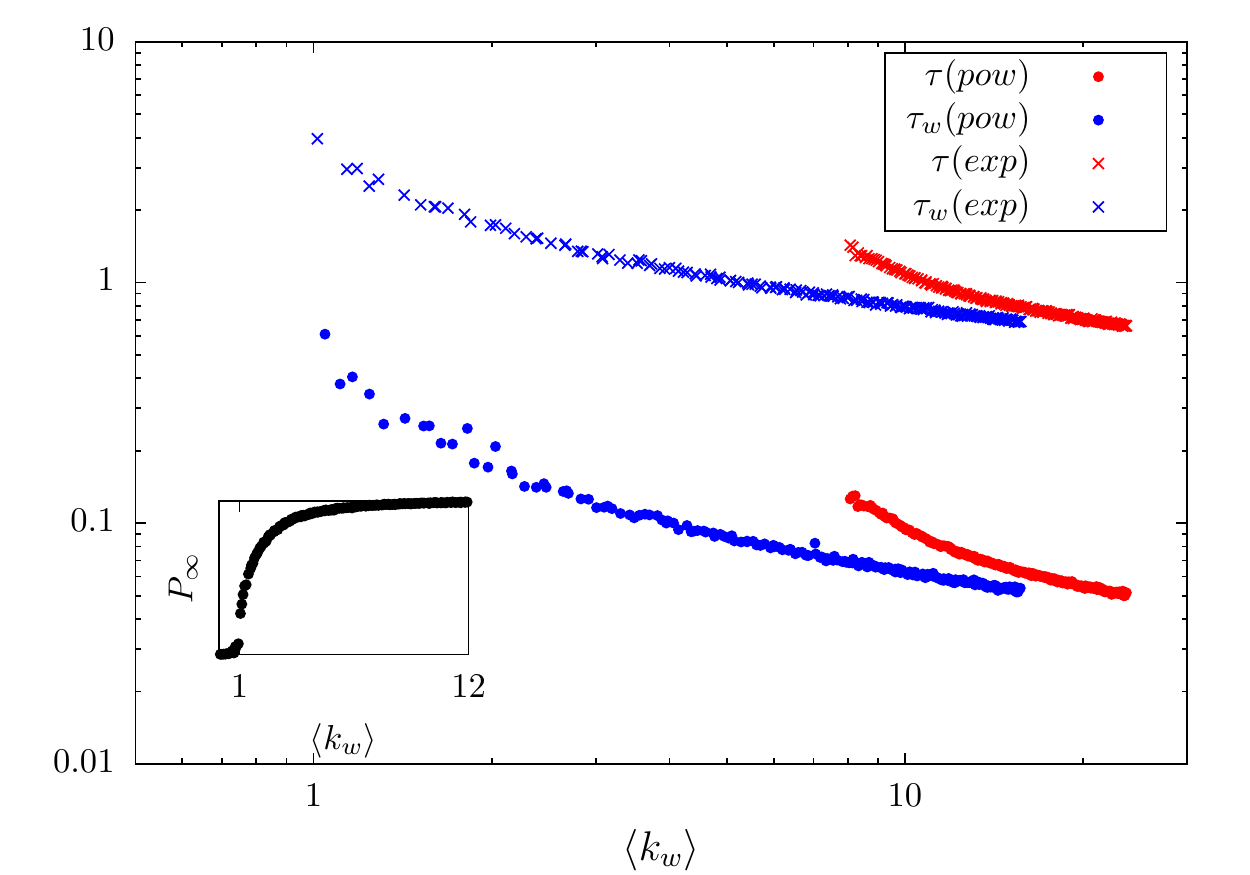}
\caption*{(d) Network of Kumpula et.al.}
\end{minipage}
\caption{Characteristic times $\tau_w$ and $\tau$ relative to average degree of white nodes $\avg{k_w}$ in the simulated networks $\GG^p_w$ and $\GG^p$ with four different initial networks $\GG$. Each network has 5000 nodes and 25000 bridges are added with both power-law ($\bullet$) and shifted exponential inter-event times ($\times$) are considered. The following models are considered: (a) Square lattice with NNN interactions; (b) Erd\H os-R\'enyi network; (c) The Barab\'asi-Albert network; (d) Network of Kumpula et.al. The data is plotted only for those values of $\avg{k}$ when the $LCC(\GG^p_w)$ exists and the results show that the process with shifted exponential inter-event times is slower on each model. Introduction of bridges decreases the characteristic time and the decrease depends on the initial topology of the network.}
\label{fig:avdeg_tau_simulation}
\end{figure}

The results are presented on the Figure~\ref{fig:avdeg_tau_simulation}. All four network models show faster spreading with bursty interaction patterns modelled by the power-law inter-event times ($\tau_w(pow)$ and $\tau(pow)$) compared to the process with shifted exponential inter-event times ($\tau_w(exp)$ and $\tau(exp)$), which was also captured in \cite{Jo_etal2014}. Introduction of bridges decreases characteristic time in all models, which agrees with the empirical results in Section~\ref{sec:results}. It is observable that the large diameter and regularity in the structure of the underlying network influences much on the speed of spreading.

\begin{figure}[h]
  \centering
  \includegraphics[width=1\linewidth]{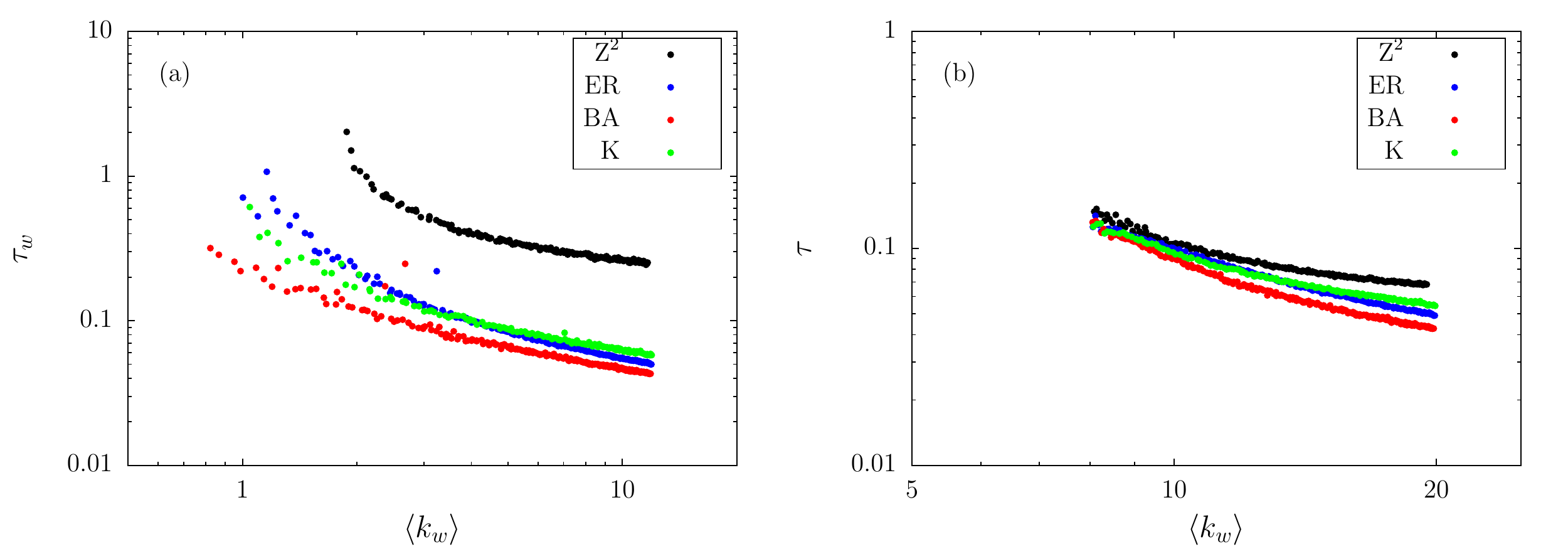}
\caption{Comparison of characteristic times $\tau_w$ and $\tau$ relative to average degree of white nodes $\avg{k_w}$ in the simulated networks $\GG^p_w$ and $\GG^p$ with four different initial networks $\GG$. Each network has 5000 nodes and 25000 bridges are added and power-law distributed inter-event times are used. The data is shown only for those values of $\avg{k}$ when the $LCC(\GG^p_w)$ exists and the results suppose that the degree distribution, or mainly the presence of large degree nodes, is a key factor governing the speed of spreading in these models. However, if these nodes belong to communities, a slowdown happens when the average degree of a network is large. Introduction of bridges decrease characteristic time and smears topological differences.}
\label{fig:avdeg_tau_kumpula}
\end{figure}

Following the observation of bursty activity patterns, on the Figure~\ref{fig:avdeg_tau_kumpula} we show the comparison between the characteristic times $\tau_w$ and $\tau$ for four considered models with power-law inter-event times. We observe that $BA$ model is the fastest in terms of characteristic time, which is explained by the presence of large degree nodes and since the percolation threshold is zero, these nodes remain in the diluted network as well. It is known that the presence of a community structure slows down spreading (see e.g. \cite{WuLiu2008}). Therefore in the model $K$ the spreading is slower than $ER$ when $\avg{k_w}$ is large. In the diluted network with small average degree $\avg{k_w}$ the $LCC$ has no apparent community structure, but since it has more large degree nodes, it overtakes $ER$ in the spreading speed. The model $Z^2$ is the slowest due to its regular structure and large diameter. The introduction of bridges smears the topology of initial networks and thus the spreading results in almost the same behaviour for all four models.

\section{Discussion}

In the current paper we have studied the role of the topology of the underlying network on the SI infection spreading process. We have observed the relation between the network structure and the characteristic time of the process. External bridges have shown to have a substantial speed up factor, thus in practice in order to define the actual speed of spreading in the social network, it is crucial to have a complete information about interactions of its members. However, the error caused by the incompleteness decays with the increase of the average degree of a network, i.e. if the average degree is sufficiently large, the speed of spreading remains almost the same regardless of the underlying network topology. Thus, the results suppose that the topology matters only if the average degree of the underlying network is small enough.

Introduction of bridges modifies the original network topology and speeds up the spreading process, which is explained by the increase of the number of possible ways the infection can spread. Bridges were shown in the Watts-Strogatz model \cite{WS} to have a major effect on the topological and dynamic properties of networks \cite{WS, Moore_Newman,Lloyd_et_al}. Here we demonstrated their importance in a system, where the original network had a broad degree distribution. 

Our observations hold for the real data experiments and for the model calculations. Empirical observations show that the presence of large degree nodes increases spreading speed, and regularity in the structure or the presence of the communities decreases it. The presented model is defined in a simple way, thus it may allow to study the observed phenomena mathematically in more detail. For the practical use it would be interesting to understand the behaviour of the spreading speed on different network models when the average degree becomes large. 

\section{Acknowledgements}

AM gratefully acknowledges financial support from the European Research Council grant ``Limits of discrete structures'', 617747 and the Grant 16-01-00499 of the Russian Foundation of Basic Research. JK acknowledges support from the H2020 FETPROACT-GSS CIMPLEX Grant No. 641191.

\section*{References}
\bibliography{literature}

\begin{thebibliography}{10}
\expandafter\ifx\csname url\endcsname\relax
  \def\url#1{\texttt{#1}}\fi
\expandafter\ifx\csname urlprefix\endcsname\relax\def\urlprefix{URL }\fi
\expandafter\ifx\csname href\endcsname\relax
  \def\href#1#2{#2} \def\path#1{#1}\fi

\bibitem{Vespignani2012}
A.~Vespignani,
  \href{http://www.nature.com/nphys/journal/v8/n1/full/nphys2160.html}{Modelling
  dynamical processes in complex socio-technical systems}, Nature Physics 8
  (2012) 32--39.
\newblock \href {http://dx.doi.org/10.1038/nphys2160}
  {\path{doi:10.1038/nphys2160}}.
\newline\urlprefix\url{http://www.nature.com/nphys/journal/v8/n1/full/nphys2160.html}

\bibitem{RevModPhys2015}
R.~Pastor-Satorras, C.~Castellano, P.~Van~Mieghem, A.~Vespignani,
  \href{http://link.aps.org/doi/10.1103/RevModPhys.87.925}{Epidemic processes
  in complex networks}, Rev. Mod. Phys. 87 (2015) 925--979.
\newblock \href {http://dx.doi.org/10.1103/RevModPhys.87.925}
  {\path{doi:10.1103/RevModPhys.87.925}}.
\newline\urlprefix\url{http://link.aps.org/doi/10.1103/RevModPhys.87.925}

\bibitem{Bansal_etal2010}
S.~Bansal, J.~Read, B.~Pourbohloul, L.~A. Meyers,
  \href{http://www.tandfonline.com/doi/abs/10.1080/17513758.2010.503376}{The
  dynamic nature of contact networks in infectious disease epidemiology},
  Journal of Biological Dynamics 4~(5) (2010) 478--489.
\newblock \href {http://dx.doi.org/10.1080/17513758.2010.503376}
  {\path{doi:10.1080/17513758.2010.503376}}.
\newline\urlprefix\url{http://www.tandfonline.com/doi/abs/10.1080/17513758.2010.503376}

\bibitem{HolmeSaramaki_PhyRep2011}
P.~Holme, J.~Saramäki,
  \href{http://www.sciencedirect.com/science/article/pii/S0370157312000841}{Temporal
  networks}, Physics Reports 519~(3) (2012) 97--125, temporal Networks.
\newblock \href
  {http://dx.doi.org/http://dx.doi.org/10.1016/j.physrep.2012.03.001}
  {\path{doi:http://dx.doi.org/10.1016/j.physrep.2012.03.001}}.
\newline\urlprefix\url{http://www.sciencedirect.com/science/article/pii/S0370157312000841}

\bibitem{Karsai_etal2011}
M.~Karsai, M.~Kivel\"a, R.~K. Pan, K.~Kaski, J.~Kert\'esz, A.-L. Barab\'asi,
  J.~Saram\"aki,
  \href{http://link.aps.org/doi/10.1103/PhysRevE.83.025102}{Small but slow
  world: How network topology and burstiness slow down spreading}, Phys. Rev. E
  83 (2011) 025102.
\newblock \href {http://dx.doi.org/10.1103/PhysRevE.83.025102}
  {\path{doi:10.1103/PhysRevE.83.025102}}.
\newline\urlprefix\url{http://link.aps.org/doi/10.1103/PhysRevE.83.025102}

\bibitem{HorvathKertesz2014}
D.~X. Horváth, J.~Kertész,
  \href{http://stacks.iop.org/1367-2630/16/i=7/a=073037}{Spreading dynamics on
  networks: the role of burstiness, topology and non-stationarity}, New Journal
  of Physics 16~(7) (2014) 073037.
\newblock \href {http://dx.doi.org/10.1088/1367-2630/16/7/073037}
  {\path{doi:10.1088/1367-2630/16/7/073037}}.
\newline\urlprefix\url{http://stacks.iop.org/1367-2630/16/i=7/a=073037}

\bibitem{Delvenne_etal2015}
J.~C. Delvenne, R.~Lambiotte, L.~E. {Correa da Rocha},
  \href{http://www.nature.com/ncomms/2015/150609/ncomms8366/full/ncomms8366.html}{Diffusion
  on networked systems is a question of time or structure}, Nature
  Communications 6.
\newblock \href {http://dx.doi.org/10.1038/ncomms8366}
  {\path{doi:10.1038/ncomms8366}}.
\newline\urlprefix\url{http://www.nature.com/ncomms/2015/150609/ncomms8366/full/ncomms8366.html}

\bibitem{Fortunato2010}
S.~Fortunato,
  \href{http://www.sciencedirect.com/science/article/pii/S0370157309002841}{Community
  detection in graphs}, Physics Reports 486~(3–5) (2010) 75--174.
\newblock \href {http://dx.doi.org/10.1016/j.physrep.2009.11.002}
  {\path{doi:10.1016/j.physrep.2009.11.002}}.
\newline\urlprefix\url{http://www.sciencedirect.com/science/article/pii/S0370157309002841}

\bibitem{Barabasi2005}
A.-L. Barabasi,
  \href{http://www.nature.com/nature/journal/v435/n7039/full/nature03459.html}{The
  origin of bursts and heavy tails in human dynamics}, Nature 435 (2005)
  207--211.
\newblock \href {http://dx.doi.org/10.1038/nature03459}
  {\path{doi:10.1038/nature03459}}.
\newline\urlprefix\url{http://www.nature.com/nature/journal/v435/n7039/full/nature03459.html}

\bibitem{IribarrenMoro2009}
J.~L. Iribarren, E.~Moro,
  \href{http://link.aps.org/doi/10.1103/PhysRevLett.103.038702}{Impact of human
  activity patterns on the dynamics of information diffusion}, Phys. Rev. Lett.
  103 (2009) 038702.
\newblock \href {http://dx.doi.org/10.1103/PhysRevLett.103.038702}
  {\path{doi:10.1103/PhysRevLett.103.038702}}.
\newline\urlprefix\url{http://link.aps.org/doi/10.1103/PhysRevLett.103.038702}

\bibitem{Miritello_etal2011}
G.~Miritello, E.~Moro, R.~Lara,
  \href{http://link.aps.org/doi/10.1103/PhysRevE.83.045102}{Dynamical strength
  of social ties in information spreading}, Phys. Rev. E 83 (2011) 045102.
\newblock \href {http://dx.doi.org/10.1103/PhysRevE.83.045102}
  {\path{doi:10.1103/PhysRevE.83.045102}}.
\newline\urlprefix\url{http://link.aps.org/doi/10.1103/PhysRevE.83.045102}

\bibitem{Vazquez_etal2007}
A.~Vazquez, B.~R\'acz, A.~Luk\'acs, A.-L. Barab\'asi,
  \href{http://link.aps.org/doi/10.1103/PhysRevLett.98.158702}{Impact of
  non-poissonian activity patterns on spreading processes}, Phys. Rev. Lett. 98
  (2007) 158702.
\newblock \href {http://dx.doi.org/10.1103/PhysRevLett.98.158702}
  {\path{doi:10.1103/PhysRevLett.98.158702}}.
\newline\urlprefix\url{http://link.aps.org/doi/10.1103/PhysRevLett.98.158702}

\bibitem{Liben-NovellKleinberg2008}
D.~Liben-Nowell, J.~Kleinberg,
  \href{http://www.pnas.org/content/105/12/4633.abstract}{Tracing information
  flow on a global scale using internet chain-letter data}, Proceedings of the
  National Academy of Sciences 105~(12) (2008) 4633--4638.
\newblock \href {http://dx.doi.org/10.1073/pnas.0708471105}
  {\path{doi:10.1073/pnas.0708471105}}.
\newline\urlprefix\url{http://www.pnas.org/content/105/12/4633.abstract}

\bibitem{Jo_etal2014}
H.-H. Jo, J.~I. Perotti, K.~Kaski, J.~Kert\'esz,
  \href{http://link.aps.org/doi/10.1103/PhysRevX.4.011041}{Analytically
  solvable model of spreading dynamics with non-poissonian processes}, Phys.
  Rev. X 4 (2014) 011041.
\newblock \href {http://dx.doi.org/10.1103/PhysRevX.4.011041}
  {\path{doi:10.1103/PhysRevX.4.011041}}.
\newline\urlprefix\url{http://link.aps.org/doi/10.1103/PhysRevX.4.011041}

\bibitem{Newman_etal2002}
M.~E.~J. Newman, S.~Forrest, J.~Balthrop,
  \href{http://link.aps.org/doi/10.1103/PhysRevE.66.035101}{Email networks and
  the spread of computer viruses}, Phys. Rev. E 66 (2002) 035101.
\newblock \href {http://dx.doi.org/10.1103/PhysRevE.66.035101}
  {\path{doi:10.1103/PhysRevE.66.035101}}.
\newline\urlprefix\url{http://link.aps.org/doi/10.1103/PhysRevE.66.035101}

\bibitem{Stehle_etal2011}
J.~Stehl{\'e}, N.~Voirin, A.~Barrat, C.~Cattuto, V.~Colizza, L.~Isella,
  C.~R{\'e}gis, J.-F. Pinton, N.~Khanafer, W.~Van~den Broeck, P.~Vanhems,
  \href{http://bmcmedicine.biomedcentral.com/articles/10.1186/1741-7015-9-87}{Simulation
  of an seir infectious disease model on the dynamic contact network of
  conference attendees}, BMC Medicine 9~(1) (2011) 1--15.
\newblock \href {http://dx.doi.org/10.1186/1741-7015-9-87}
  {\path{doi:10.1186/1741-7015-9-87}}.
\newline\urlprefix\url{http://bmcmedicine.biomedcentral.com/articles/10.1186/1741-7015-9-87}

\bibitem{Blondel_etal2015}
V.~D. Blondel, A.~Decuyper, G.~Krings,
  \href{http://dx.doi.org/10.1140/epjds/s13688-015-0046-0}{{A survey of results
  on mobile phone datasets analysis}}, EPJ Data Science 4~(1).
\newblock \href {http://dx.doi.org/10.1140/epjds/s13688-015-0046-0}
  {\path{doi:10.1140/epjds/s13688-015-0046-0}}.
\newline\urlprefix\url{http://dx.doi.org/10.1140/epjds/s13688-015-0046-0}

\bibitem{MedvedevPete2016}
A.~Medvedev, The role of the topology in non-poissonian spreading dynamics on
  temporal networks, Ph.D. thesis, Central European University (2016).

\bibitem{Pan_etal2011}
R.~K. Pan, M.~Kivel\"a, J.~Saram\"aki, K.~Kaski, J.~Kert\'esz,
  \href{http://link.aps.org/doi/10.1103/PhysRevE.83.046112}{Using explosive
  percolation in analysis of real-world networks}, Phys. Rev. E 83 (2011)
  046112.
\newblock \href {http://dx.doi.org/10.1103/PhysRevE.83.046112}
  {\path{doi:10.1103/PhysRevE.83.046112}}.
\newline\urlprefix\url{http://link.aps.org/doi/10.1103/PhysRevE.83.046112}

\bibitem{BarabasiAlbert1999}
A.-L. Barab{\'a}si, R.~Albert,
  \href{http://science.sciencemag.org/content/286/5439/509}{Emergence of
  scaling in random networks}, Science 286~(5439) (1999) 509--512.
\newblock \href {http://dx.doi.org/10.1126/science.286.5439.509}
  {\path{doi:10.1126/science.286.5439.509}}.
\newline\urlprefix\url{http://science.sciencemag.org/content/286/5439/509}

\bibitem{ErdosRenyi1959}
P.~Erd\"{o}s, A.~R\'{e}nyi, On random graphs i, Publicationes Mathematicae
  Debrecen 6 (1959) 290.

\bibitem{KSKK07}
J.~M. Kumpula, J.-P. Onnela, J.~Saram\"aki, K.~Kaski, J.~Kert\'esz,
  \href{http://link.aps.org/doi/10.1103/PhysRevLett.99.228701}{Emergence of
  communities in weighted networks}, Phys. Rev. Lett. 99 (2007) 228701.
\newblock \href {http://dx.doi.org/10.1103/PhysRevLett.99.228701}
  {\path{doi:10.1103/PhysRevLett.99.228701}}.
\newline\urlprefix\url{http://link.aps.org/doi/10.1103/PhysRevLett.99.228701}

\bibitem{WuLiu2008}
X.~Wu, Z.~Liu,
  \href{http://www.sciencedirect.com/science/article/pii/S0378437107010254}{How
  community structure influences epidemic spread in social networks}, Physica
  A: Statistical Mechanics and its Applications 387~(2–3) (2008) 623--630.
\newblock \href {http://dx.doi.org/10.1016/j.physa.2007.09.039}
  {\path{doi:10.1016/j.physa.2007.09.039}}.
\newline\urlprefix\url{http://www.sciencedirect.com/science/article/pii/S0378437107010254}

\bibitem{WS}
D.~J. Watts, S.~Strogatz,
  \href{http://www.nature.com/nature/journal/v393/n6684/full/393440a0.html}{Collective
  dynamics of small world networks}, Nature 393 (1998) 440--442.
\newblock \href {http://dx.doi.org/10.1038/30918} {\path{doi:10.1038/30918}}.
\newline\urlprefix\url{http://www.nature.com/nature/journal/v393/n6684/full/393440a0.html}

\bibitem{Moore_Newman}
C.~Moore, M.~E.~J. Newman,
  \href{http://journals.aps.org/pre/abstract/10.1103/PhysRevE.61.5678}{Epidemics
  and percolation in small-world networks}, Phys. Rev. E 61 (2000) 5678--5682.
\newblock \href {http://dx.doi.org/10.1103/PhysRevE.61.5678}
  {\path{doi:10.1103/PhysRevE.61.5678}}.
\newline\urlprefix\url{http://journals.aps.org/pre/abstract/10.1103/PhysRevE.61.5678}

\bibitem{Lloyd_et_al}
A.~L. Lloyd, S.~Valeika, A.~Cintron-Arias,
  \href{http://www.ams.org/books/conm/410/}{Infection dynamics on small-world
  networks}, in: A.~B. Gumel, C.~Castillo-Chavez, D.~Clemence (Eds.),
  Mathematical Studies on Human Disease Dynamics: Emerging Paradigms and
  Challenges, AMS, 2006, pp. 222--247.
\newblock \href {http://dx.doi.org/10.1090/conm/410}
  {\path{doi:10.1090/conm/410}}.
\newline\urlprefix\url{http://www.ams.org/books/conm/410/}

\end{thebibliography}

\end{document}